# Size dependent thermoelectric properties of silicon nanowires

Lihong Shi[1], Donglai Yao[1], Gang Zhang[2, a)] and Baowen Li[1,3]

[1]*Centre for Computational Science and Engineering and Department of Physics, National University of Singapore, 117542 Singapore*

[2]*Institute of Microelectronics, A*STAR, 11 Science Park Road, Singapore Science Park II, Singapore 117685, Singapore*

[3]*NUS Graduate School for Integrative Sciences and Engineering, 117597 Singapore*

**Abstract**

By using first-principles tight-binding electronic structure calculation and Boltzmann transport equation, we investigate the size dependence of thermoelectric properties of silicon nanowires (SiNWs). With cross section area increasing, the electrical conductivity increases slowly, while the Seebeck coefficient reduces remarkably. This leads to a quick reduction of cooling power factor with diameter. Moreover, the figure of merit also decreases with transverse size. Our results demonstrate that in thermoelectric application, NW with small diameter is preferred. We also predict that isotopic doping can increase the value of *ZT* significantly. With 50% $^{29}$Si doping ($^{28}$Si$_{0.5}$$^{29}$Si$_{0.5}$ NW), the *ZT* can be increased by 31%.



a) Email: zhangg@ime.a-star.edu.sg




Thermoelectric material plays an important role in solving the energy crisis. The cooling efficiency is given by figure of merit, $ZT = \frac{S^2 \times \sigma}{\kappa}T$, here $S$ is the Seebeck coefficient, $\sigma$ is the electrical conductivity, $T$ is the absolute temperature, and $\kappa$ is the thermal conductivity. $\kappa = \kappa_e + \kappa_p$, where $\kappa_e$ and $\kappa_p$ are the electron and phonon (lattice vibration) contribution to the thermal conductivity, respectively. Silicon nanowire (SiNW) has attracted broad interests in recent years due to its ideal interface compatibility with Si-based electronic technology and the potential applications. [1, 2] It has been demonstrated that the thermal conductivity of SiNW can be 100 times smaller than that of bulk silicon. [3-6] .Vo *et al.* [7] have studied the impacts of growth direction and doping on the thermoelectric figure of merit of SiNWs by using *ab initio* electronic structure calculations. In addition to the growth direction and carrier concentration, it has been demonstrated that the electronic structure of SiNW also depends on the transverse size remarkably [8-11]. Moreover, it is well established that isotope doping is an efficient way to reduce the thermal conductivity. [12-14] In this letter, we will investigate the impacts of size and isotope concentration on thermoelectric property of SiNW.

We focus on SiNWs oriented along the [110] direction with rectangular cross section shape, which can be fabricated experimentally. [10, 11, 15] Here the cross section of SiNW is from 1 nm$^2$ to about 18 nm$^2$. The surface dangling bonds are terminated with hydrogen atoms. A supercell approach is adopted where each wire is periodically repeated along the growth direction [110]. The size of the supercells in the transverse plane is large enough (>15 Å from surface to surface). In this letter, the density functional derived tight-binding method (DFTB) [16, 17] is used. The structural relaxation is performed using a conjugate gradient method. The atomic force tolerance of $3\times10^{-4}$ eV/Å is applied. Self-consistent charge



tolerance is $10^{-5}$ au. The accuracy of DFTB method in SiNW band structure calculation has been demonstrated. [8, 18]

The electrical conductivity σ, the thermal conductivity due to electrons $\kappa_e$, and the Seebeck coefficient *S*, are obtained from the electronic structure with the solution of 1-Dimensional Boltzmann transport equation as:

$$\sigma = \Lambda^{(0)}$$
$$\kappa_e = \frac{1}{e^2 T}[\Lambda^{(2)} - \Lambda^{(1)}(\Lambda^{(0)})^{-1}\Lambda^{(1)}]$$
$$S = \frac{1}{eT}(\Lambda^{(0)})^{-1}\Lambda^{(1)} \quad (1)$$
$$\Lambda^{(n)} = e^2 \tau \frac{2}{m^*} \sum_{E_k} \Delta E \left[ \frac{\beta \exp(\beta(E_k - \mu))}{(1 + \exp(\beta(E_k - \mu)))^2} \right] D(E_k) E_k (E_k - \mu)^n$$

Here *e* is the charge of carriers, *T* is the temperature, $E_k$ is the electron energy, τ is the relaxation time, m* is the effective mass of the charge carrier, μ is the electron chemical potential and $D(E_k)$ is the density of states. In this work, we obtain the value of relaxation time τ by fitting the calculated mobility to measured electrical conductivity data of SiNW. As the lack of experimental electrical conductivity data on diameter dependence, here we use the experimental data of SiNW with fixed diameter of 48 nm (n=1.7×10$^{19}$ cm$^{-3}$, σ=588 (Ωcm)$^{-1}$, from Ref. 5), and obtain $\tau = 4.3 \times 10^{-16}$ s. Then, we use the dependence relation [19] between the mobility and carrier concentration in bulk silicon to calculate the carrier concentration dependent relaxation time. For example, $\tau = 1.7 \times 10^{-15}$ s for n=1.0×10$^{17}$ cm$^{-3}$, and $\tau = 9.7 \times 10^{-16}$ for n=5.0×10$^{17}$ cm$^{-3}$. As *ZT* of n-doped SiNWs is considerably larger than that of their p-doped counterparts, we only study n-doped wires in this letter. The carrier concentration is defined as: $n = \int D(E - \mu) \times f(E - \mu) \times dE$, where *f(E)* is the Fermi distribution function.

Figure 1(a) and 1(b) show the size effects on σ and *S* with different electron



concentration. σ increases slightly as the diameter increases, while the Seebeck coefficient $S$ decreases remarkably. The size dependence arises from quantum confinement effect on the electronic band structure. Figure 2 shows the density of states (DOS) of intrinsic SiNWs. It is obvious that the larger the dimension of the wire the smaller the band gap. However, as the electron band gap converges quickly with transverse dimension increases, [8-11] σ only has an obvious size dependence in very small size (less than 5 nm$^2$). In contrast, Seebeck coefficient $S$ decreases with increasing of size remarkably. Besides the electronic band gap, Seebeck coefficient S also depends on the detailed band structure, in which narrow DOS distribution is preferred. [20, 21] In SiNW, the large numbers of electronic stats in narrow energy ranges can lead to large S. With the transverse dimension increases, the sharp DOS peaks widen and reduces S. The increase in transverse dimension has two effects on band structure: reduce the band gap; and widen the sharp DOS peaks, both have negative impacts on Seebeck coefficient. So the Seebeck coefficient decreases quickly with transverse size increasing. In addition to the transverse size dependence, both $S$ and σ depend on carrier concentration. As shown in Figure 1(c) and (d), $S$ decreases as the carrier concentration increases, while σ increases as more carriers are available to transport charge.

In thermoelectric application, the power factor $P$ ($P = S^2 \times \sigma$), is an important factor influencing the thermoelectric performance directly. Figure 3 (a) shows the power factor versus carrier concentration. There is an optimal carrier concentration $N_{Max}$ yielding the maximum attainable value of $P_{Max}$. As SiNW area increases, the $P_{Max}$ decreases (figure 3(b)), and $N_{Max}$ shifts to lower carrier concentration (figure 3(c)). The slow increase of σ is offset by obvious decreasing in $S$ ($P \sim S^2$), as a result the power factor reduces with size increasing. In Figure 3 (b), at small size, the small increases of cross section area can induce large



reduction on power factor. For example, with the cross section area increases from 1.1 to 4.7 nm$^2$, the power factor decreases with about 4300 μW/m-K$^2$. In contrast, the power factor versus cross section area curves are almost flat when cross section area increases from 14.1 to 17.8 nm$^2$.

The power factor (*P*) is related to the cooling power density (*PD*) of a thermoelectric cooler. In the practical application, the maximal cooling power density is given by [22]: $PD_{MAX} = \frac{1}{2}\frac{S^2\sigma}{L}T^2$, here *T* is the environment temperature around the SiNW, and *L* is the length of the SiNW. In our analysis, we consider SiNW with length of 1μm. Figure 3(d) shows the size dependence of maximal cooling power density versus the cross section area at room temperature. Even for thick SiNW with cross section area of 17.8 nm$^2$, the maximal cooling power density, $6.4\times10^3$ W/cm$^2$, is about six times larger than that of SiGeC/Si superlattice coolers [23], ten times larger than that of Si/Si$_x$Ge$_{1-x}$ cooler [24], and six hundred times larger than that of commercial TE module. [25]

The figure of merit, *ZT* is another important factor for thermoelectric materials. In the calculation of *ZT*, both electron and phonon contribute to the total thermal conductivity. Figure 4(a) shows the dependence of electron thermal conductivity (calculated from Eq. 1) on carrier concentration. It is clear that κ$_e$ increases with carrier concentration and SiNW with larger diameter is with higher κ$_e$. And κ$_p$ of SiNW increases with diameter increases remarkably until the diameter is larger than about hundreds nms. [26, 27].Thus κ increases with transverse dimension. Combine the size dependence of the power factor as shown in figure 3(b), we can conclude that *ZT* will decrease when the NW diameter increases. However, at present, there is no consensus in the literature about quantitative expression of size dependence of lattice (phonon) thermal conductivity in thin SiNWs, which limits the



study of size effect on *ZT* quantitatively.

Besides diameter impact, the isotope doping is an important method to tune the thermal conductivity of nano materials. [12] In this letter, we focus on the isotope doping effect on *ZT* of SiNW ($^{28}Si_{1-x}^{29}Si_x$ NWs) with fixed cross section area of 2.3 nm$^2$. From molecular dynamics calculation, [12] $\kappa_p$ of pure $^{28}Si$ NW is 1.49 W/m-K. In Figure 4(a) it is obvious that with the carrier concentration we studied here, is much larger than. Using our calculated S, $\sigma$, $\kappa_e$ from Eq. 1, and $\kappa_p$ value from Ref. 12 (calculated by using non-equilibrium molecular dynamics method, Stillinger-Weber (SW) potential and Nosé-Hoover /Langevin heat bath, more details in Ref 12), the dependence of *ZT* on carrier concentration is shown in figure 4(b). Similar to the power factor, there exists one optimal carrier concentration $N_{Max}$ yielding the maximum attainable value of $ZT_{Max}$. Moreover, the isotope doping concentration changes the value of $ZT_{Max}$ remarkably. Figure 4(c) and 4(d) show the dependence of $ZT_{Max}$ and $N_{Max}$ on concentration of $^{29}Si$ (x). The $ZT_{Max}$ values increase with $^{29}Si$ concentration, reach a maximum and then decreases. In the case of $^{28}Si_{0.8}^{29}Si_{0.2}$ NW, namely, 20% $^{29}Si$, its $ZT_{Max}$ increases with 15% from that of pure $^{28}Si$ NW. And with 50% $^{29}Si$ doping ($^{28}Si_{0.5}^{29}Si_{0.5}$ NW), the $ZT_{Max}$ can increase with 31%. The carrier concentration $N_{Max}$ to attain $ZT_{Max}$ is independent on the isotope concentration, as the $^{29}Si$ doping only changes the lattice vibration and has no effect on the electronic structures in our calculations.

In summary, we have investigated the size dependence of thermoelectric properties of SiNWs. With cross section area increases, the electrical conductivity increases slowly while the Seebeck coefficient reduces remarkably which leads to a quick reduction of cooling power factor. Moreover, the figure of merit, *ZT* also decreases with increasing of transverse size. Our results have demonstrated that in thermoelectric application, NW with small



diameter is preferred. In addition to the size effect, we have also predicted that the isotopic doping can increase the value of *ZT*. With 50% $^{29}$Si doping ($^{28}$Si$_{0.5}$$^{29}$Si$_{0.5}$ NW), the *ZT* can be increased by 31%, from 0.81 to 1.06.

**Acknowledgment.**

The work is supported in part by grant R-144-000-203-112 from Ministry of Education of Republic of Singapore and grant R-144-000-222-646 from National University of Singapore, and Grant No. R-144-000-243-597 from the US Air Force.

**References**

[1] Y. Cui, Q. Q. Wei, H. K. Park, C. M. Lieber, Science, **293**, 1289, (2001).

[2] Y. Cui, C. M. Lieber, Science, **291**, 851, (2001).

[3] S. G. Volz and G. Chen, Appl. Phys. Lett. **75**, 2056 (1999).

[4] D. Li, Y. Wu, P. Kim, L. Shi, P. Yang, and A. Majumdar, Appl. Phys. Lett. **83**, 2934 (2003).

[5] A. I. Hochbaum, R. Chen, R. D. Delgado, W. Liang, E. C. Garnett, M. Najarian, A. Majumdar, and P. Yang, Nature **451**, 163 (2008).

[6] A. I. Boukai, Y. Bunimovich, J. T. Kheli, J-K Yu, W. A. Goddard III, and J. R. Heath, Nature **451**, 168 (2008).

[7] T. Vo, A. J. Williamson, V. Lordi, and G. Galli, Nano Lett. **8**, 1111 (2008).

[8] D. Yao, G. Zhang, and B. Li, Nano Lett., **8**, 4557 (2008).

[9] D. D. D. Ma, C. S. Lee, F. C. K. Au, S. Y. Tong, S. T. Lee, Science, **299**, 1874, (2003).

[10] T. Vo, A. J. Williamson, G. Galli, Physical Review B, **74**, 045116, (2006).

[11] M. Nolan, S. O'Callaghan, G. Fagas, and J. C. Greer, Nano Letters, **7**, 34 (2007).

[12] N. Yang, G. Zhang, and B. Li, Nano Lett. **8**, 276 (2008).

[13] G. Zhang and B. Li, Journal of Chemical Physics, **123**, 114714, (2005).

[14] C. W. Chang, D. Okawa, H. Garcia, A. Majumdar, A. Zettl, Phys. Rev. Lett. **101**, 075903, (2008).

[15] R. Q. Zhang, Y. Lifshitz, D. D. D. Ma, Y. L. Zhao, T. Frauenheim, S. T. Lee, and S. Y. Tong, Journal of Chemical Physics, **123**, 144703 (2005).

[16] M. Elstner, D. Porezag, G. Jungnickel, J. Elsner, M. Haugk, T. Frauenheim, S. Suhai, G. Seifert, Phys. Rev. B **58**, 7260 (1998).

[17] T. Frauenheim, G. Seifert, M. Elstner, Z. Hajnal, G. Jungnickel, D. Porezag, S. Suhai, R. Scholz, Phys. Status Solidi B **217**, 41 (2000).

[18] D. Yao, G. Zhang, and B. Li, Appl. Phys. Lett., **94**, 113113, (2009).

[19] C. Jacoboni, C. Canali, G. Ottaviani, and A. A. Quaranta, Solid-State Electronics, **20**, 77, (1977).

[20] T. E. Humphrey, H. Linke, Phys. Rev. Lett., **94**, 096601, (2005).

[21] R. Y. Wang, J. P. Feser, J-S Lee, D. V. Talapin, R. Segalman, and A. Mujumdar, Nano Lett., **8**, 2283, (2008).

[22] G. Zhang, Q. X. Zhang, C. T. Bui, G. Q. Lo, and B. Li, Appl. Phys. Lett., **94**, 213108 (2009).

[23] X. Fan, G. Zeng, C. Labounty, J. E. Bowers, E. Croke, C. C. Ahn, S. Huxtable, A. Majumdar, and A. Shakouri, Appl. Phys. Lett. **78**, 1580, (2001).

[24] Y. Zhang, J. Christofferson, A. Shakouri, G. Zeng, J. E. Bowers, and E. T. Croke, IEEE Transactions on Components and Packaging Technologies, **29**, 395, (2006).

[25] Thermoelectrics Handbook Macro to Nano, edited by D. M. Rowe, Taylor & Francis Group, (2006).

[26] P. K. Schelling, S. R. Phillpot, and P. Keblinski, Phys. Rev. B, **65**, 144306, (2002).

[27] L. H. Liang and B. Li, Phys. Rev. B, **73**, 153303, (2006).

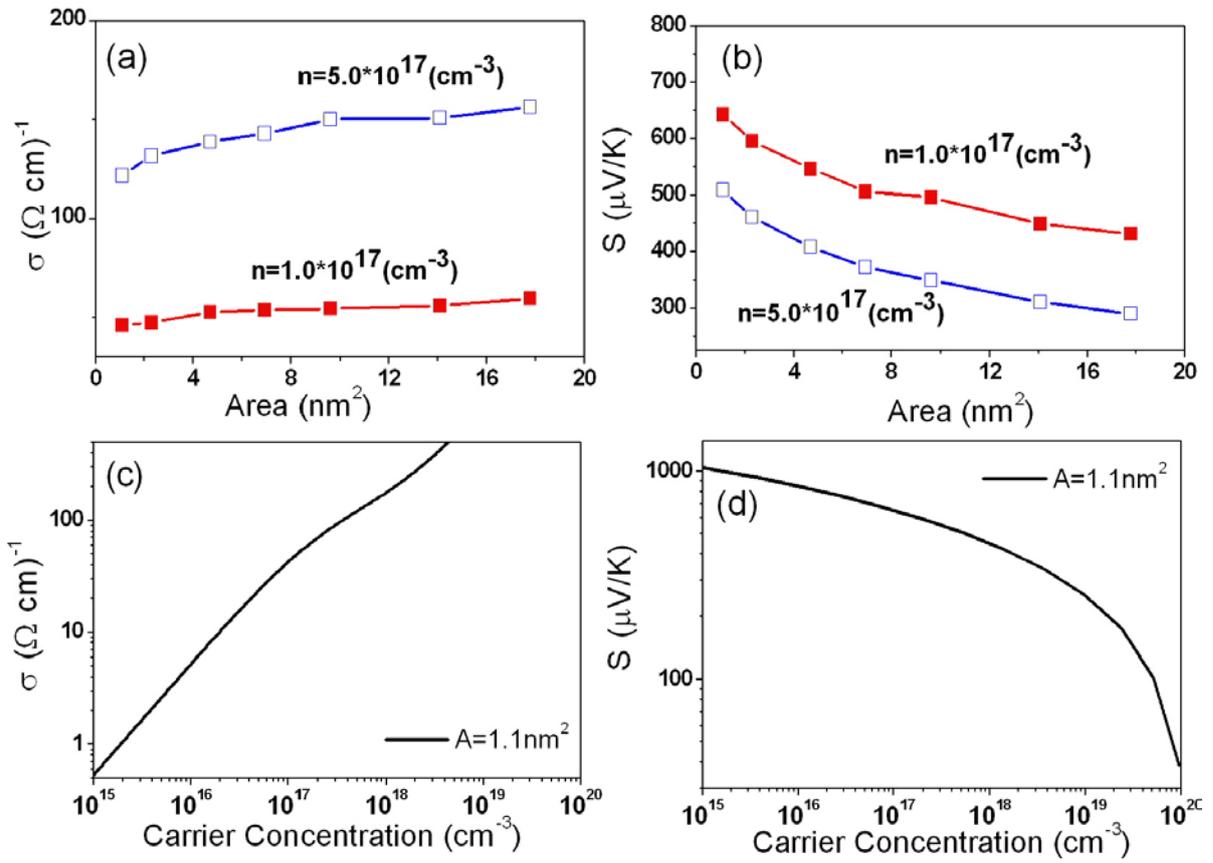

Figure 1. (a) σ vs cross sectional area with different carrier concentration. (b) *S* vs cross sectional area with different carrier concentration. (c) σ vs carrier concentration with fixed cross section area of 1.1 nm$^2$. (d) *S* vs carrier concentration with fixed area of 1.1 nm$^2$.

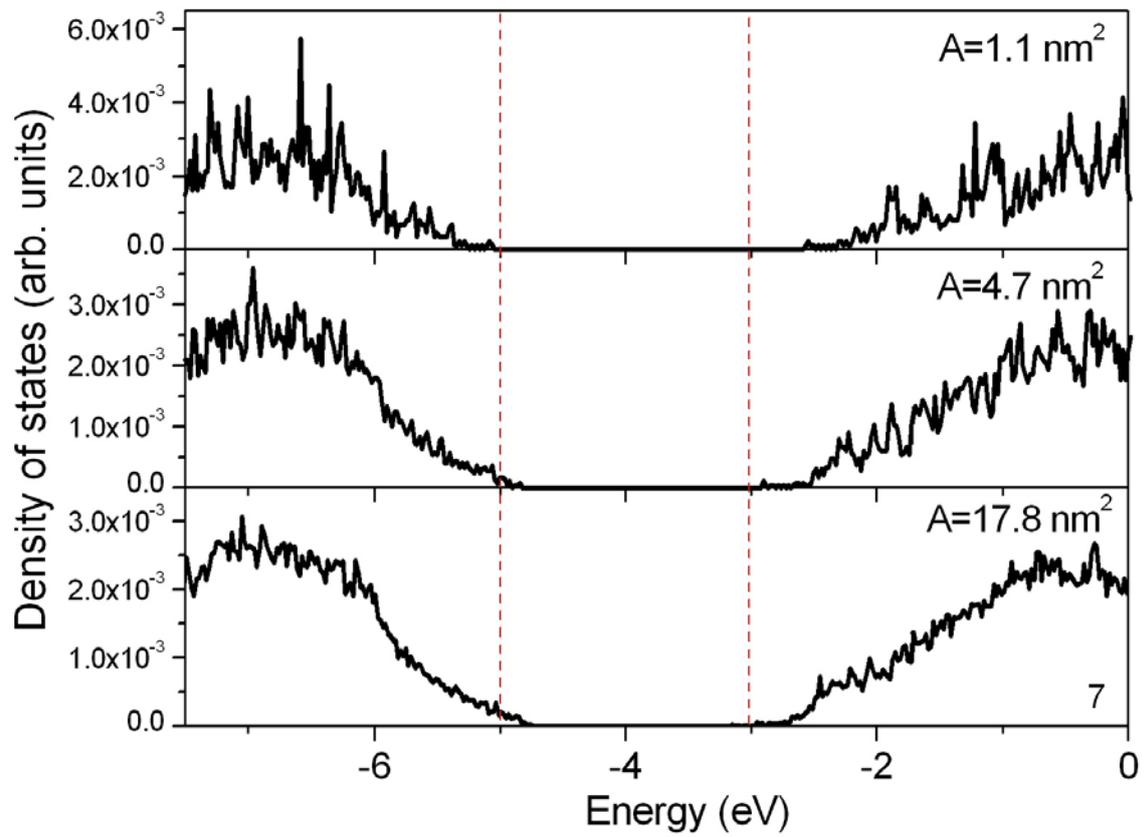

Figure 2. DOS for SiNWs with three different transverse dimensions from 1.1 to 17.8 nm$^2$. The red dotted lines are drawn to guide the eyes.

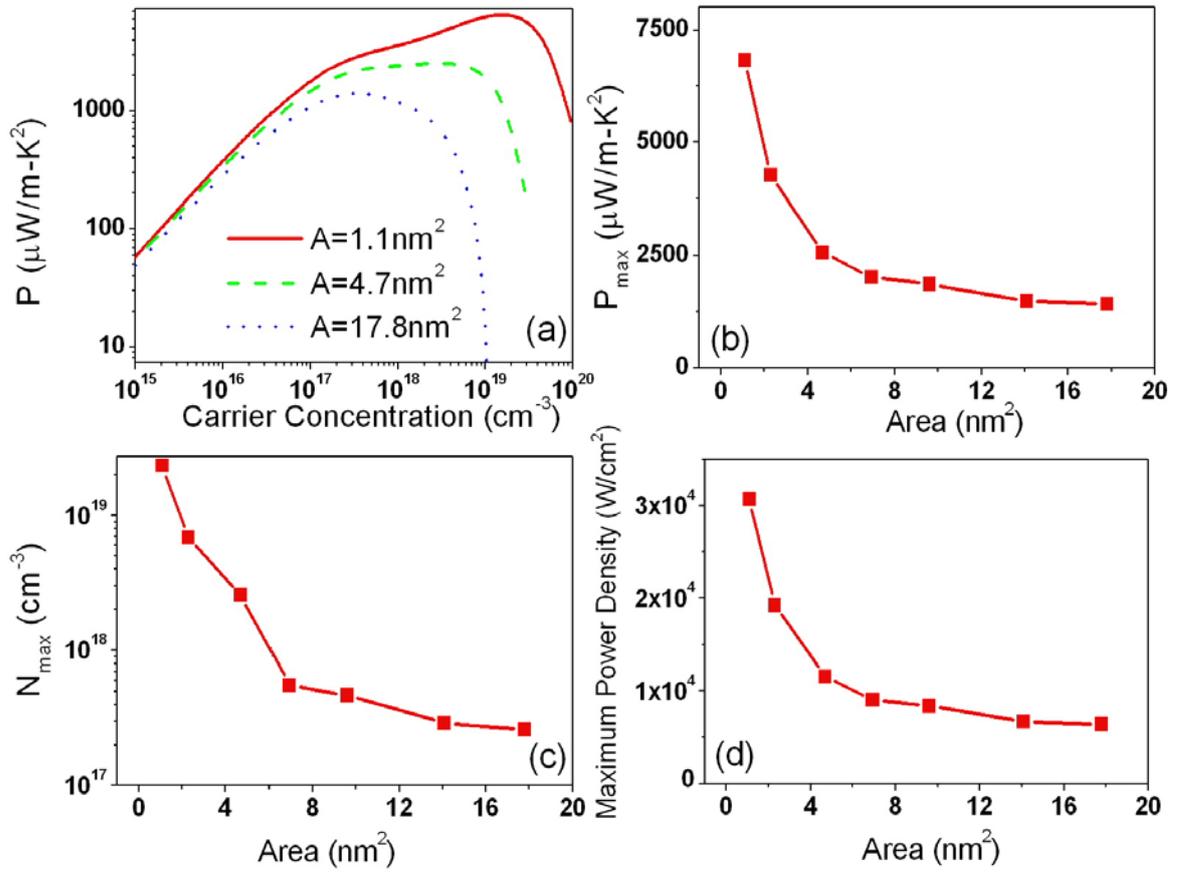

Figure 3. (a) Thermal power factor of SiNW vs carrier concentration with three different transverse dimensions. (b) Maximum power factor vs cross sectional area. (c) $N_{Max}$ vs cross sectional area. (d) Size dependence of the maximum room temperature cooling power density of SiNW with length of 1μm.

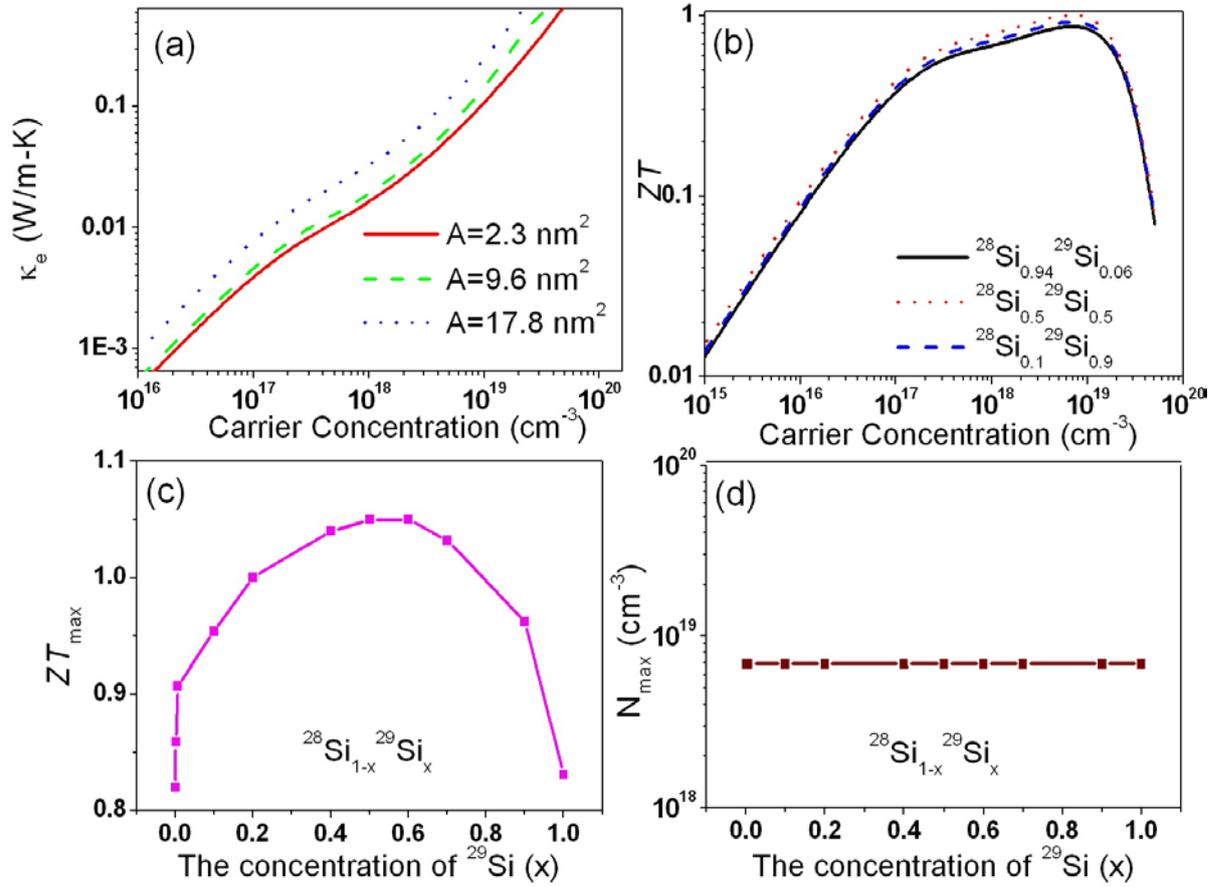

Figure 4. (a) Thermal conductivity due to electrons vs carrier concentration for SiNWs with different transverse dimensions. (b) $ZT$ vs carrier concentration for different isotope-doped SiNWs ($^{28}Si_{1-x}{}^{29}Si_x$ NWs) with fixed cross section area of 2.3 nm$^2$. (c) $ZT_{Max}$ vs the concentration of $^{29}Si$ atom. (d) $N_{Max}$ vs the concentration of $^{29}Si$ doping atom.